\documentclass[12pt,preprint]{aastex}

\shorttitle{SMBH growth with ALMA}
\shortauthors{Kawakatu et al.}

\begin{document}

\title{Exploring Supermassive Black Hole Growth with ALMA}


\author{Nozomu Kawakatu\altaffilmark{1}}
\affil{National Astronomical Observatory of Japan, 2-21-1 Osawa, 
Mitaka, Tokyo 151-8588, Japan}

\author{Paola Andreani}
\affil{European Southern Observatory, Karl Schwarzschild strasse 2, D-
85748 Garching, Germany}

\author{Gian Luigi Granato}
\affil{INAF-Ossevatorio Astronomica di Padova. Vicolo dell'Ossevatorio 5, I-35100, Padova, Italy\\SISSA/ISAS, via Beirut 2-4, 34014, Trieste, Italy}

\author{Luigi Danese}
\affil{SISSA/ISAS, via Beirut 2-4, 34014, Trieste, Italy\\INAF-Ossevatorio Astronomica di Padova. Vicolo dell'Ossevatorio 5, I-35100, Padova, Italy}


\altaffiltext{1}{kawakatu@th.nao.ac.jp}


%%%%%%%%%%%%%%%%%%%%%%%%%%%%%%%%%%%
% (2)Abstract  & Subject Headings %
%%%%%%%%%%%%%%%%%%%%%%%%%%%%%%%%%%%
\begin{abstract}

Massive tori with $\approx 10^{8-9}M_{\odot}$ are predicted to extend on 
$\sim $100 pc scale around the centre of elliptical galaxy progenitors 
by a model of a supermassive black hole (SMBH) growth coeval to the spheroidal 
population of the host galaxy.
Direct detection of such massive tori would cast light on a key physical
condition that allows the rapid growth of SMBHs and the appearance of 
QSOs at high redshift. 
For this reason, we examine the detectability of such structures
at substantial redshift with the Atacama Large Millimeter Array (ALMA). 
We propose that submillimeter galaxies (SMGs) are the best
targets to test our predictions. In order to assess the observational
feasibility, we estimate the expected number counts of SMGs with
massive tori and check the detectability with the ALMA instrument,
the unique facility which can resolve the central region of high
redshift objects.
Our work shows that ALMA will be able to resolve and detect high-$J$ 
($J >$ 4) CO emissions from $\sim$100 pc scale extended massive tori up to 
$z\approx2$. 
Observations of lensed SMGs will yield excellent spatial resolution, 
allowing  even to resolve their massive tori at higher redshift.
We discuss further the detectability of the
HCN molecule, as a better tracer of the high density gas expected in
such tori. 
The final goal of these kind of observations is to pinpoint possible 
physical mechanisms that storage in the very central galactic regions 
very large amount gas on timescale of several 10$^{8}$ yr.

\end{abstract}

%\newpage
%%%%%%%%%%%%%%%%%%%%%%%%%%%%%
% (3)TEXT & Acknowledgments %
%%%%%%%%%%%%%%%%%%%%%%%%%%%%%
\section{Introduction}

The energy emitted by active galactic nuclei (AGNs) is commonly
ascribed to accretion onto a supermassive black hole (SMBH).
Recent high-resolution observations of galactic centers indicate
the presence of a SMBH whose mass correlates with the mass, the
velocity dispersion, and the luminosity of the galactic bulge
(e.g., Kormendy \& Richstone 1995; Richstone et al. 1998; Magorrian et al. 1998; Laor 1998; Ferrarese \& Merritt 2000; Tremaine et al. 2002; 
McLure \& Dunlop 2001; 2002; Marconi \& Hunt 2003). QSO hosts are mostly
luminous and well evolved early-type galaxies (e.g., McLeod \&
Rieke 1995a; Bahcall et al. 1997; McLure, Dunlop \& Kukula 2000;
Dunlop et al. 2003). 
%At high redshifts they are associated to the large
%metallicity, dusty environments (e.g., Hamann \& Ferland 1993; Hamann
%\& Ferland 1999; Freudling et al 2003; Dietrich et al. 2003;
%Nagao et al. 2006, Andreani et al, 1999; Maiolino et al. 2004).
Taken together, these observations indicates that the formation of
a SMBH, a bulge and a QSO is closely related to each other. While
it is still not clear which is the main mechanism for the
formation of a SMBH, many studies suggest that a BH growth by
gas accretion processes plays an important if not dominant role
(Salucci et al. 1999; Fabian \& Iwasawa 1999; Yu \& Tremaine 2002; 
Hosokawa 2002; Marconi et al. 2004; Shankar et al. 2004). 
%The discovery of powerful QSOs at z$>$6 (Fan et al.\ 2001) implies the
%formation of big SMBHs in less than 1 Gyr.

Kawakatu, Umemura \& Mori 2003 (hereafter KUM03) and Granato et
al. 2004 (hereafter G04) proposed a physical model for the
co-evolution of QSO BHs and the spheroidal component of galaxies.
The formation of these latter is computed with a semi-analytical
technique, which includes the effects of stellar
and QSO feedback, and model outcomes have been more widely compared 
to observations of several galaxy
populations (G04; Silva et al. 2005; Granato et al.
2006; Lapi et al. 2006). In these models the fueling of
the central BH in forming spheroids takes place in two steps: (i)
a low-angular momentum gas {\it reservoir} is formed at a rate
essentially proportional to the star formation rate on a time
scale of a few $10^{8}\,{\rm yr}$, then (ii) the
reservoir gas is accreted onto the central BH at a rate of the order of
the Eddington limit. Since in the earlier phases of a SMBH growth
the former rate largely exceeds the latter, a massive reservoir
accumulates; most of its mass is accreted onto the SMBH only at 
the last {\it e}-folding times ($t_e\simeq 4\times 10^7$ yr) and powers 
an enormous energy release, which stops both the star formation 
and the growth of the reservoir. 
The massive reservoir is likely to be a geometrically thick 
massive disk (we call it massive tori) because of some residual 
angular momentum. We will comment on the passage from the 
massive reservoir to the massive torus in details later.
A plausible physical mechanism responsible for step (i) is
the radiation drag (Umemura 2001; Kawakatu \& Umemura 2002), thus
hereafter we refer them as the radiation drag model. It should be
however kept in mind that any other mechanism producing a similar
rate of production of massive tori, would not
alter the essence of these computations.

The scenario proposed by KUM03 and G04 establishes a well defined
sequence connecting various populations of massive spheroidal 
galaxies: (i) vigorous and rapidly
dust-enshrouded star formation activity, during which a central
SMBH grows, and explaining the sub-mm galaxy population (SMG), 
which are examples of high-$z$ ultraluminous infrared galaxies 
(ULIRGs);
(ii) QSO phase halting subsequent star formation and (iii)
essentially passive evolution of stellar populations, going
through an Extremely Red Object (ERO) phase (see Fig. 1; For
details Fig. 2 and Fig. 3 in KUM03 and Fig.3 in G04).

Here we suggest that the massive torus, before accreting onto the
SMBH (time-scale is $\approx 10^{8-9}\,{\rm yr}$), 
forms a massive torus of the kind 
envisaged by unified models of AGNs (e.g., Antonucci 
1993; Granato \& Danese 1994; Urry \& Padovani 1995). 
Thus the radiation drag model predicts the
existence of a massive torus with $\approx 10^{8-9}M_{\odot}$ in the early
phase of SMBH and spheroid growth (stage (i) and also the beginning 
of stage (ii), see Fig. 1). 
The impact of the direct detection of massive tori 
is to reveal whether the large mass ratio of 
a torus to a black hole, $M_{\rm torus}/M_{\rm BH}\gg 1$, 
is a key physical condition which triggers the rapid growth of SMBHs 
and the appearance of QSOs. This issue will be discussed in details later.  
Hence, we investigate the required spatial resolution and expected CO
 and HCN emission lines from massive tori. We focus our investigation
on the high-$z$ universe ($z> 1$)
where the peak of QSO activity occurs (around $z \simeq 2$) and mild
AGN activity has already been detected in high-$z$ SMGs (Alexander et
al. 2003, 2005a, \& 2005b) in very good agreement with the
G04 predictions (Granato et al.\ 2006). In particular, we
here investigate the detectability of massive tori at $z>1$ with the
next generation millimeter interferometer, ALMA (Atacama Large
Millimeter Array). 
The rest of this paper is organized as follows. In $\S 2$, we briefly 
review the formation of massive tori in the early phase of the BH growth 
and set up the model for massive tori.
In $\S 3$, we first estimate the expected spatial resolution and
expected CO and HCN fluxes from massive tori. Secondly, we discuss
the best targets with massive tori and then show the number
count of candidates with massive tori. Finally, 
we mention the importance of the velocity mapping on massive tori, 
and the impact on the detection of them for SMBH growth.
Our discussions and conclusions 
are given by $\S 4$. Through this paper, we adopt the cosmology 
indicated by {\it WMAP} data (Bennett et al. 2003; Spergel et al. 2006), 
i.e., a flat Universe with the Hubble parameter $H_{0}=70\,{\rm
km}\,{\rm s}^{-1}\,{\rm Mpc}^{-1}$, $\Omega_{\rm M}=0.27$ and
$\Omega_{\Lambda}=0.73$.

%\section{Model for massive tori}
%On the basis of the radiation drag model, we consider $\sim$ 100 pc
%scale dynamically stable massive torus
%with a uniform density (see \S 2.3 in G04). 
%In addition, this model predicts that the mass of a massive torus $M_{\rm %torus}$ is equal to QSO BHs with $\approx 10^{8-9}M_{\odot}$ (see Fig. 1).

\section{Model}
\subsection{Mass accretion due to radiation drag}

We review a physical model of the formation of a SMBH in forming spheroids, 
based on the radiation drag-driven mass accretion. 
The radiation drag, driving the mass accretion, originates in the relativistic effect in absorption and subsequent re-emission of the radiation. This effect is naturally involved in relativistic radiation hydrodynamic equations (Fukue, Umemura, \& Mineshige 1997). 
The angular momentum transfer in radiation hydrodynamics is 
given by the azimuthal equation of motion in cylindrical coordinates, 
\begin{equation}
\frac{1}{r}\frac{d(rv_{\rm \phi})}{dt}=\frac{\chi_{\rm d}}{c}[F^{\rm \phi}-
(E+P^{\rm \phi \phi})v_{\rm \phi}], \label{ldot}
\end{equation} 
where $E$ is the radiation energy density, $F^{\rm \phi}$ 
is the radiation flux, $P^{\rm \phi\phi}$ is the radiation stress tensor,
and $\chi_{\rm d}$ is the mass extinction coefficient which is given 
by $\chi_{\rm d}=n_{\rm d}\sigma_{\rm d}/\rho_{gas}$ with the number density of dust grains $n_{\rm d}$, the dust cross-section $\sigma_{\rm d}$ and the gas density $\rho_{\rm gas}$.
By solving the radiative transfer including dust opacity, Kawakatu \& Umemura (2002) evaluated the radiative quantities in a clumpy medium, $E$, $F^{\rm \phi}$, and $P^{\rm \phi\phi}$, and thereby obtained the total angular momentum loss rate. 
Then, these authors estimated the total mass of the dusty ISM accreted on to a central massive {\it reservoir}, $M_{\rm res}$. 
By using the relation, $\dot{M}_{\rm gas}/M_{\rm gas}=-\dot{J}/J$, 
where $J$ and $M_{\rm gas}$ are the total angular momentum and gas of ISM. 
Then,  the {\it reservoir} mass is assessed as     
\begin{equation}
M_{\rm res}(t)=\int_{0}^{t}\dot{M}_{\rm gas}dt 
=-\int_{0}^{t}M_{\rm gas}\frac{\dot{J}}{J}dt.
\end{equation}
In the optically-thick regime of the radiation drag,
$-M_{\rm gas}\dot{J}/{J}=L_{\rm sph} (t)/c^{2}$, where $L_{\rm sph}(t)$ is
the total luminosity of spheroids.
The radiation drag efficiency depends on the optical depth $\tau$ 
in proportion to $(1-e^{-\tau})$ (Umemura 2001). 
Thus, the total mass of {\it reservoir} can be expressed by 
\begin{equation}
M_{\rm res}(t)=\eta_{\rm drag}\int^{t}_{0}
\frac{L_{{\rm sph}}(t^{\prime})}
{c^2}\left(1-e^{-\tau (t^{\prime})}\right) dt^{\prime},
\end{equation}
where $\tau(t)$ is the optical depth of spheroids measured 
from the center. 
Here, we estimate the evolution of $\tau(t)$ by using 
an evolutionary spectral synthesis code 'PEGASE' 
(Fioc \& Rocca-Volmerange 1997). 
The efficiency $\eta_{\rm drag}$ is found to be maximally 0.34 
(Kawakatu $\&$ Umemura 2002).

Moreover, it is worth keeping in mind the ratio of {\it reservoir} mass to spheroidal mass (see Umemura 2001 in details). 
The radiation energy emitted by a main sequence star is 
$0.14\epsilon$ times the rest mass energy of the star $m_{*}c^{2}$, 
where $\epsilon$ is the energy conversion efficiency of nuclear 
fusion from hydrogen to helium, which is 0.007. 
Thus, the total luminosity of spheroids is estimated to be 
\begin{equation}
L_{\rm sph}(t)\simeq 0.14\epsilon S_{\rm sph}(t)c^{2}, 
\end{equation}
where $S_{\rm sph}(t)$ is the star formation rate in the spheroidal system.
By substituting this in equation (3), 
the {\it reservoir} mass to spheroidal mass is given by 
\begin{equation}
\frac{M_{\rm res}(t)}{M_{\rm sph}(t)}\simeq 2\times 10^{-3}, 
\end{equation}
which is comparable to the observed black hole-to-bulge mass ratio 
in nearby galaxies. 
Here $M_{\rm sph}=\int^{t}_{0}S_{\rm sph}(t^{\prime})dt^{\prime}$. 

\subsection{Formation of a massive torus}
In $\S 2.1$, we introduced the {\it reservoir} mass in the context 
of the radiation drag-induced mass accretion. 
However, the massive {\it reservoir} does not evolve into 
a SMBH directly, because the radiation drag can not remove 
the angular momentum thoroughly (Sato et al. 2004). 
Thus, some residual angular momentum would terminate the 
radial contraction. 
In the massive {\it reservoir}, the viscosity is expected 
to work effectively because the viscous timescale is shrinked 
by the radiation drag (Mineshige et al. 1998). 
Thus, the massive {\it reservoir} is likely to be a massive 
self-gravitating viscous disk. 
In such a massive disk the extreme star formation 
occurs in the outer part of a massive torus via the gravitational instability, 
since Toomre $Q$ parameter is less than unity, where $\kappa c_{\rm s}/\pi 
G\Sigma_{\rm g}$ for the epicycle frequency $\kappa$, the sound velocity 
$c_{\rm s}$ and the surface density $\Sigma_{\rm g}$. 
If this is the case, the massive {\it reservoir} will become a 
{\it geometrically thick} massive disk (hereafter we call it a massive torus: 
$M_{\rm torus}=M_{\rm res}$), 
by supporting the vertical thickness by the energy feedback from supernovae 
(Wada \& Norman 2002) and radiation pressure from the starburst 
(Ohsuga \& Umemura 2001; Thompson et al. 2005; Watabe \& Umemura 2005).

As for the BH growth, we here assume that the mass accretion driven by 
the viscosity onto the BH horizon is determined by an order of 
Eddington rate, $\dot{M}_{\rm Edd}\equiv L_{\rm Edd}/c^{2}\simeq0.2M_{\odot}{\rm yr}^{-1}(M_{\rm BH}/10^{8}M_{\odot})$, where $L_{\rm Edd}$ is the Eddington 
luminosity. 
As mentioned in $\S 2.1$, the mass accretion rate from a host galaxy to a galactic center via radiation drag is $\dot{M}_{\rm drag}\approx L_{\rm sph}/c^{2}\simeq
0.1(L_{\rm sph}/10^{12}L_{\odot})M_{\odot}{\rm yr}^{-1}$. 
According to KUM03 (also G04), the starburst phase is closely related to 
the formation of a QSO. 
During the first stage, vigorous star formation occurs, while a SMBH 
is still growing ($M_{\rm BH}\approx 10^{6}-10^{7}M_{\odot}$). 
In this phase, the existence of the massive torus 
with $M_{\rm torus}/M_{\rm BH}\gg 1$ can be predicted, since 
$\dot{M}_{\rm drag}$ largely exceeds $\dot{M}_{\rm Edd}$. 
At the end of starburst phase, the SMBH has grown and radiates 
at maximum luminosity (QSO phase). In the QSO phase, $M_{\rm BH}$ 
approaches $M_{\rm torus}$ and thus $M_{\rm torus}/M_{\rm BH}\simeq 1$. 
After the QSO phase, the AGN luminosity decreases rapidly, 
since almost all of the matter in massive tori has fallen onto the 
central BH, namely QSO would evolve into low luminosity AGNs (LLAGNs).
Hence, $M_{\rm torus}/M_{\rm BH}$ can be expected to be much smaller 
than unity for spheroidal galaxies with LLAGNs. 
This is consistent with observation which 
elliptical galaxies and galactic bulges have no gas. 
Therefore, it can be concluded that the evolution of the mass ratio 
($M_{\rm torus}/M_{\rm BH}$) reflects the history of the SMBH growth.

\subsection{Physical parameters of massive tori}

The radiation drag mechanism, as mentioned $\S\S$ 2.1 and 2.2, 
predicts the presence of massive tori endowed with 
a mass $M_{\rm torus}\simeq 10^{8}-10^{9}$ M$_{\odot}$ equal to 
the final mass of BHs powering QSOs (see Fig.1). 
Typical length scales of such tori can be expressed by the accretion 
radius of the $^{``}$BH plus massive torus" system ($r_{\rm a}$).  
The accretion radius is given by $r_{\rm a}=G(M_{\rm BH}+M_{\rm torus})
/\sigma_{\rm sph}^{2}$, where $\sigma_{\rm sph}$ is the velocity dispersion 
of spheroids. 
Following the virial theorem for spheroids ($\sigma_{\rm sph}^{2}
=GM_{\rm sph}/r_{\rm b}$), the radius can be written as 
\begin{equation}
r_{\rm a}=\left(\frac{M_{\rm BH}+M_{\rm torus}}{M_{\rm sph}}\right)r_{\rm b}
\simeq \left(\frac{M_{\rm torus}}{M_{\rm sph}}\right)r_{\rm b}, 
\end{equation}
where $r_{\rm b}$ is the effective radius of spheroids. 
Here we consider the case of $M_{\rm BH}\ll M_{\rm torus}$ since 
we focus on the rapid growth phase (see stage (i) in Fig. 1).
By substituting equation (5) in equation (6), the $r_{\rm a}$ can 
be estimated by $r_{\rm a}\simeq 20$ pc $(r_{\rm b}/10\,{\rm kpc})$. 
In the present paper, the size of a massive torus is defined as 
$r_{\rm torus}=A r_{\rm a}$. 
The factor $A$ can be inferred by high resolution observations of tori 
in nearby AGN (e.g., Jaffe et al. 1993; van der Marel et al. 1998; Davis et al. 2006) and turns out to be at least $A \approx 5$. 
Recent IR and X-ray observations support $\sim$100 pc scale extended massive tori (e.g., McLeod \& Rieke 1995b; Maiolino et al. 1995; Maiolino \& Rieke 1995; 
Granato et al. 1997; Malkan, Gorjian \& Tran 1998) rather than the compact massive tori.
However, $r_{\rm torus}$ (or $A$) is physically determined by the quantity 
of residual angular momentum removed by the radiation drag process. 
In order to reveal this issue, we should explore the sophisticated 
numerical simulations, which are our future study.
Thus, a compact massive torus can not be ruled out physically. 
In this paper, we examine two cases; one is a compact torus with 
$r_{\rm torus}=20$ pc ($A$=1), the other is an extended torus with 
$r_{\rm torus} =100$ pc ($A$=5). 

Concerning the rotation law and the surface density profile 
in massive tori, we assume $v_{\phi}(r)=v_{\phi,0}(r/r_{\rm torus})^{-\alpha}$ 
and 
$\Sigma_{\rm g}(r)=\Sigma_{\rm g,0}(r/r_{\rm torus})^{-\beta}$, 
where $v_{\phi,0}$ and $\Sigma_{\rm g,0}$ are the quantities 
at $r_{\rm torus}$. 
$\Sigma_{\rm g,0}$ is related to the torus mass through 
$M_{\rm torus}=\int^{r_{\rm torus}}_{r_{\rm in}} 2\pi r\Sigma_{\rm g}(r){\rm d}r$,
where $r_{\rm in}$ is the sublimation radius 
for graphite grains. The size $r_{\rm in}$ is expressed as $r_{\rm in}=1.3\, L_{\rm UV, 46}^{0.5}T_{1500}^{-2.8}\,{\rm pc}$, where $L_{\rm UV, 46}$ is the AGN UV luminosity in unit of $10^{46}{\rm erg}\,{\rm s}^{-1}$, and $T_{1500}$ is the grain sublimation temperature in unit of 1500 K (Barvainis 1987). 
Recently, Suganuma (2006) measured the inner edge of dusty torus $r_{\rm in}$ 
by the reverberation mapping technique, and thier observational results 
strongly support that $r_{\rm in}$ is determined by the dust 
sublimation radius. 
Although the size $r_{\rm in}$ depends on the size of dust grains and 
dust composition (e.g., Laor \& Draine 1993), these does not affect on 
the expected line flux significantly as far as $r_{\rm in}\ll r_{\rm torus}$ .
Thus, we set up $r_{\rm in}$=1 pc in the present paper. 
The rotation velocity, $v_{\phi,0}=A^{-\alpha}v_{\phi}(r_{\rm a})$, 
can be determined from $v_{\phi}(r_{\rm a})=\sigma_{\rm sph}$. 
Since $\alpha$ and $\beta$ are unknown free parameters, we will 
calculate the CO flux density corresponding to the following 
parameter set ($\alpha=-1, 0, 0.5$ and $\beta=0, 1, 2$). 

Provided that the scale height ($h_{\rm torus}$) of massive tori 
is $\sim$0.3$\times r_{\rm torus}$ (e.g., Wada \& Norman 2002), 
the average number density and column density of molecular hydrogen are 
$n({\rm H}_{2})\approx 10^{3}\,{\rm cm}^{-3}(M_{\rm gas}/10^{8}M_{\odot})
(r_{\rm torus}/100\,{\rm pc})^{-3}$ and $N({\rm H}_{2})\approx 10^{23}{\rm cm}^{-2}(M_{\rm gas}/10^{8}M_{\odot})(r_{\rm torus}/100\,{\rm pc})^{-2}$, respectively. 
Thus massive tori are optically thick and thermalized at 
CO lines. This happens to be for any molecular clouds with
hydrogen column density larger than $N({\rm H}_{2}) > 10^{21}\,{\rm cm}^{-2}$ (e.g., Tielens \& Hollenbach 1985). For $h_{\rm torus}$, we expect that the 
opening angle is almost zero in the quite extreme conditions we are envisaging.
However, the expected CO flux density does not depend on $h_{\rm torus}$
as long as the massive tori are optically thick for CO lines.
 
As for the gas temperature $T_{\rm g}$ of the optically thick massive torus, 
Ohsuga \& Umemura (2001) analyzed it by using one-dimensional radiation hydrodynamic equations, coupling with ionization processes 
and thermal processes. 
They found that dust cooling is effective and thus 
the gas temperature $T_{\rm g}$ is $\approx 100\, {\rm K}$. Thus, we assume $T_{\rm g}=100\, {\rm K}$ in this paper. 
This is also supported by the recent numerical results that the massive 
tori would be dominated by the cold ($T_{\rm g} \sim  100$ K) and 
dense ($n({\rm H}_{2})\approx 10^{2}{\rm cm}^{-3}$) molecular gas (e.g., Wada \& Tomisaka 2005).
Finally, we note that the CO luminosity and the CO-to-H$_{2}$ conversion 
factor are a function of the metallicity (Radford, Solomon \& Downes 1991) and the far-UV flux from AGNs and/or massive stars lower of a few 
magnitudes for $n({\rm H}_{2}) > 10^{2}\,{\rm cm}^{-3}$ (e.g., Tielens \& Hollenbach 1985; Mochizuki \& Nakagawa 2000).

%On the mass accretion
%process onto a central BH, almost all gas would fall onto a BH from
%$\sim$ 100 pc during $\approx 10^{7-8}$ yr if the angular momentum
%transfer via the turbulent viscosity continues to work effectively in
%massive tori (e.g., Duschl et al. 2000; Burkert \& Silk 2001; Wada \& Norm%an 1999, 2002; also see Wada 2004). 
%Here we should keep in mind that it is not
%clear whether the turbulent viscosity lasts for the AGN life timescale.
%For hydrogen molecular gas in gravitationally bound, the mass of the hydro%gen molecule $M_{H_{2}}$ is traced by the CO luminosity $X_{\rm CO}=M(H_{2%})/L'_{\rm CO}(1,0)$. In the Milky Way, the $X_{\rm CO}$-parameter is $4.6%M_{\odot}/{\rm K}\,{\rm km}{\rm s}^{-1}\,{\rm pc}^{2}$. However, the $\alp%ha$-parameter could affect the CO abundance (Sakamoto 1996) and Arimoto et% al. (1996) found smaller $\alpha$ in the central regions than in outer re%gions of galaxy. In fact, low values of $\alpha$ have been suggested for t%he central regions of galaxies (e.g., Regan 2000).***quantity***
%Moreover, the metallicity of central parts in NLS1s are super-solar (e.g.%, Wills et al. 1999). Thus, the $X_{\rm CO}$-parameter could be even small%er in the central dense interstellar matter of NLS1s than the Galactic loc%al value by a factor of two or as much as ten.
%In this paper, we assume $X_{\rm CO}=1M_{\odot}/{\rm K}\,{\rm km}s^{-1}\,{%\rm pc}^{2}$.

\section{Results}
\subsection{Required spatial resolution}

Fig. 2 shows the spatial resolution required to detect 100 pc
scale extended massive tori. 
%This is given by
%\begin{equation}
%\Theta (^{\prime\prime}) = \frac{r_{\rm torus}}{d_{\rm
%A}}\frac{180\times 3600}{\pi},
%\end{equation}
%where $d_{\rm A}$ is the angular distance. 
A resolution of $\approx 0.01^{\prime\prime}$ is necessary, 
well beyond the capabilities of present-day mm and sub-mm interferometers
(typically $\approx 0.5^{\prime\prime}$). By converse, the angular resolution of ALMA from $86-720$ GHz with $\Theta (^{\prime\prime})\simeq 0.005 (^{\prime\prime}) (\nu_{\rm obs}/600\, {\rm GHz})^{-1}$ \footnote{This comes from the ALMA webpage (http://www.eso.org/projects/alma/science/bin/sensitivity.html). 
See note 3 in this webpage.}, 
is sufficient to resolve high-$J$ ($J>4$) molecular emissions from 100 pc scale extended massive tori, 
where $\nu_{\rm obs}$ is the observed frequency. 
On the other hands, 20 pc scale compact massive tori cannot be resolved unless the objects are 
gravitationally lensed (e.g., Downes and Solomon 2003). The lensing effect magnifies not only the intrinsic CO line luminosity, 
but also the scale of emitting region.
If we assume the magnified CO image by an ellipse, the magnification factor is $\mu = a/r_{\rm torus}$, where $a=D_{\rm A}\Theta_{\rm obs}/2$ is the apparent semi-major axies of the magnified image, where $D_{\rm A}$ and $\Theta_{\rm obs}$ are the angular distance and the observed angular length of magnified image, respectively. 
Thus, it is possible to achieve the super-resolution which can be 
$\mu$ times as high as the instrumental resolution, e.g., $\mu =2-50$ 
for lensed SMGs (Solomon \& Vanden Bout 2005). 
As seen in Fig. 2, for lensed objects we can resolve lower $J$-CO lines from 
massive tori and/or study massive tori at higher redshift.

\subsection{Expected flux of the CO molecule}

%For gravitationally bound hydrogen molecular, the mass of the
%hydrogen molecule $M({\rm H_{2}})$ is traced by the CO luminosity.
Since most of molecular clouds in massive tori are in cold and dense 
($\S 2$), it is expected that CO molecular lines 
are good tracers of molecular gas mass. 
In this paper, we express the CO line luminosity $l'_{\rm CO}(J, J-1)$ 
in unit of ${\rm K}\,{\rm km}\,{\rm s}^{-1}$ as the product 
of the velocity-integrated source brightness temperature $T_{\rm b}
\Delta v_{\rm rest}$, where $\Delta v_{\rm rest}$ is the rest-frame 
line width. 
Then, the CO line luminosity in the arbitrary transitional level 
from $J$ to $J-1$ is given by $l'_{\rm CO}(J, J-1)=T_{\rm b}
\Delta v_{\rm rest}$. 
Thus, the CO line luminosity ratio for two lines, $R_{J,J-1}=l'_{\rm CO}
(J, J-1)/l'_{\rm CO}(1, 0)$, in the same source is equal to 
the ratio of the intrinsic brightness temperature. 
%The CO luminosity is given by $L'_{\rm CO}(J, J-1)=(R_{J, J-1}/X_{\rm
%CO})M({\rm H_{2}})$ where $X_{\rm CO}=M({\rm H_{2}})/L'_{\rm
%CO}(1,0)$ and the line luminosity ratio 
%$R_{J,J-1}=L'_{\rm CO}(J, J-1)/L'_{\rm CO}(1, 0)$.
Since we deal with the case that CO lines are optically thick and 
thermalized up to $J=6$, we set $R_{J,J-1}=1$. 
Note that the gas temperature can be warmer than in normal
galaxies due both to the intense radiation field via the huge
star-forming activity and to the accretion onto their
central growing BH. 
%Additionally, $T_{\rm g}$ may increase due to the higher background 
%temperature from the cosmic microwave background radiation at $z > 1$ 
%(e.g., Silk \& Spaans 1997; Combes, Maoli, \& Omont 1999). 
A flat distribution of Rayleigh-Jeans brightness temperature up to $J$=6 has been reported in the starburst galaxies 
(e.g., Kawabe et al. 1999).
%dominated by a warm and high density gas ($T_{\rm g}\approx 50-100\, {\rm %K}$ and $n ({\rm H}_{2}) 
%> 10^{4}{\rm cm }^{-3}$) dominates (e.g., Kawabe et al. 1999).

The line flux density $f_{\rm CO}(J,J-1)$ in unit of ${\rm Jy\, {\rm pc}^{-2}}$ corresponding to the
CO line luminosity $l'_{\rm CO}(J, J-1)$ can be derived.
Using H$_{2}$ mass-to-CO line luminosity relation $l'_{\rm CO}(J,J-1)(r)
=R_{J,J-1}\Sigma({\rm H_{2}})(r)/X_{\rm CO}$, 
the corresponding received CO flux density $f_{\rm CO}(J,J-1)$ is 
\begin{equation}
f_{\rm CO}(J, J-1) (r)=\frac{2\kappa}{c^{2}}\frac{l'_{\rm CO}(J, J-1)(r)\nu_{\rm rest}^{2}(J,J-1)(1+z)}{\Delta v_{\rm rest}(r)}\frac{cos(i)}{D_{\rm L}^{2}},
\end{equation}
where 
$\kappa$ is the Boltzman constant, $i$ is inclination angle of the massive tori, the $f_{\rm CO}(J,J-1)$ is the observed flux density in ${\rm Jy\, {\rm pc}^{-2}}$, $\Sigma({\rm H}_{2})(r)=\Sigma_{\rm g}(r)$, $\Delta v_{\rm rest}(r)=v_{\phi}(r)$, 
$\nu_{\rm rest}(J,J-1)=115\,J\,{\rm GHz}$ is the rest frequency of
the transition and $D_{\rm L}$ is the luminosity distance. 
Here $X_{\rm CO}=0.8M_{\odot} ({\rm K}\, {\rm km}\, {\rm s}^{-1})^{-1}$ 
is a typical value adopted for ULIRGs and SMGs (Downes \& Solomon 1998; Solomon \& Vanden Bout 2005). 
This value is comparable to $X_{\rm CO}\simeq 0.5M_{\odot} 
({\rm K}\, {\rm km}\, {\rm s}^{-1})^{-1}$, which was calculated by three-dimensional, non-local thermal equilibrium radiative transfer calculations for high-$J$ CO lines (e.g., $J=3$ and $J=4$) emitted from inhomogeneous dusty tori 
(Wada \& Tomisaka 2005).

The integrated CO flux density (Jy) of massive tori over their surface 
can be calculated by 
$S_{\rm CO}(J, J-1)=\int^{r_{\rm torus}}_{r_{\rm in}} 
2\pi rf_{\rm CO}(J, J-1)(r){\rm d}r$.
Then,
\begin{eqnarray}
S_{\rm CO}(J, J-1)&\approx &10^{3}{\rm Jy}\,J^{2}\,H(\alpha=0.5, \beta=1)
\left(\frac{X_{\rm CO}}{0.8}\right)^{-1}\left(\frac{R_{J,J-1}}{1.0}\right) \nonumber \\
&&\left(\frac{M_{\rm torus}}{3\times 10^{8}M_{\odot}}\right) 
\left(\frac{\Delta v_{\rm rest,0}}{100{\rm km/s}}\right)^{-1}\frac{1+z}{D_{\rm L}^{2}},
\end{eqnarray}
where $\Delta v_{\rm rest,0}=100\,{\rm km/s}\,(A/5)^{-(\alpha/0.5)}(M_{\rm torus}/3\times 10^{8}M_{\odot})^{0.5}(r_{\rm a}/20\,{\rm pc})^{-0.5}$, 
a inclination angle being equal to $45$ degree, 
and $H(\alpha, \beta)$ is the function of $\alpha$ and $\beta$. 
We summarize the value of $H(\alpha, \beta)$ in Table 1.

%In eq.(8), we estimate $S_{\rm CO}(J,J-1)$ in the case of 
%, $A=5$ (extendend massive tori), 
%$\alpha=0.5$ and $\beta=1$. 
%% <Consistency check>
%As for the consistency check of the value derived by eq. (2), we evaluate expec%ted CO flux density using $S_{\rm CO}(J, J-1)\approx (2\nu_{\rm rest}^{2}kT_{\r%m b}/c^{2})(\pi r_{\rm torus}^{2}/D_{L}^{2})(1+z)$ with the brightness temperat%ure $T_{\rm b}(=T_{\rm g}=100)$ K. Then, we find that the estimated $S_{\rm CO}%$ equals to one derived by eq. (8) within factor 1.5. 
%
%%
%where $M({\rm H}_{2})=M_{\rm torus}$, $\Delta v_{\rm rest}\simeq 100\,{\rm km/s%}\,(M_{\rm torus}/3\times10^{8}M_{\odot})^{0.5}(r_{\rm torus}/100\,{\rm pc})^{-%0.5}$ and normalized all dependencies to convenient values. 
%$X_{\rm CO}=M({\rm H_{2}})/L'_{\rm CO}(1,0)=0.8M_{\odot} 
%({\rm K}\, {\rm km}\, {\rm s}^{-1}\, {\rm pc}^{2})^{-1}$ 
%is a typical value adopted for ULIRGs and SMGs 
%(Downes \& Solomon 1998; Solomon \& Vanden Bout 2005). 
%Here we should emphasize that our simple estimation of CO flux density 
%has been confirmed by three-dimensional, non-local thermal equilibrium radiativ%e transfer calculations for high-$J$ CO lines (e.g., $J=3$ and $J=4$) emitted f%rom inhomogeneous dusty tori (Wada \& Tomisaka 2005).

Fig. 3 shows the expected CO emission from 100 pc scale extended 
massive tori as a function of redshift for a fiducial case 
($M_{\rm torus}=3\times 10^{8}M_{\odot}$, $A=5$, $\alpha=0.5$, and $\beta=1$). 
The five lines denote the CO flux density $S_{\rm CO}(J,J-1)$ at the
transition level ($J$, $J-1$). At the left side of filled circles, 
each line can be resolved with ALMA (see Fig.1). The CO lines without filled circles will not be resolved. 
The horizontal line shows the ALMA detection limit (50 antennas) 
0.5 mJy requested $5\sigma$ in 25
km/s channel in 12 hours of on-source integration time
\footnote{Note that the ALMA detection limit for lines depends on the
observed bands as follows; 0.31 mJy for band 3 (86-116 GHz), 0.42
mJy for band 4 (125-163 GHz), 0.39 mJy for band 5 (163-211 GHz),
0.46 mJy for band 6 (211-275 GHz) and 0.50 mJy for band 7 (275-370
GHz), wchih can be obtained by the ALMA sensitivity calculator as mentioned on 
the ALMA webpage (http://www.eso.org/projects/alma/science/bin/sensitivity.html). Here we take the greatest value (0.50 mJy).}.
The high-$J$ CO lines ($J > 4$) can be detectable and resolvable 
up to $z=2$ with ALMA. 
Again it is worth recalling that for gravitational lensed objects 
we can detect lower $J$-CO lines and/or we can study higher redshift 
objects, because the observed CO flux density increases $\mu$ times 
with respect to that estimated in eq.(8). 

Finally, we discuss the dependence of $A$ (or $r_{\rm torus}$), 
$\alpha$ and $\beta$ for $S_{\rm CO}(J,J-1)$. 
First, we investigate the effect of the function $H(\alpha,\beta)$. 
As seen in Table 1, it is found that $H(\alpha,\beta)$ is order of unity, 
except for the case of $\alpha=-1$ and $\beta=2$. 
In the case of $\alpha=-1$ and $\beta=2$, 
the expected CO flux density $S_{\rm CO}(J,J-1)$ 
has larger value compared with other cases, 
because of a high gas concentration and slower 
rotation velocity in the inner region of massive tori. 
As mentioned in $\S 2.3$, $\Delta v_{\rm rest,0}$ is linked to the 
size of the torus. 
Thus, the emissivity $S_{\rm CO}(J,J-1)$ also depends on the difference of 
$r_{\rm torus}$ (or $A$).  
Then, the CO flux density from compact massive tori ($A=1$, $\alpha=0.5$ and $0 \le \beta < 2$) is $\sim 3$ times as small as fiducial case, 
because $\Delta v_{\rm rest,0}$ of compact massive tori is larger than 
that in the fiducial case. 
Thus, the detection of compact massive tori with ALMA would be 
more difficult than the extended massive tori, taking into account 
of the spatial resolution ($\S 3.1$). 
Thirdly, the variation of $\alpha$ also affect on $S_{\rm CO}(J,J-1)$ 
for the extended massive tori ($A=5$ and $0\le \beta < 2$). 
From eq.(8), $S_{\rm CO}(J,J-1)$ of extended massive tori with 
rigid rotation ($\alpha=-1$) and flat rotation ($\alpha=0$) are 
about 0.1 and 0.5 times smaller that in the fiducial case, respectively.
From this, we find that it will be  difficult to detect 
the extended massive tori with ALMA if they rotate rigidly. 
To sum up in $\S 3.1$ and $3.2$, we conclude that 100 pc extended 
massive tori can be resolved and detected with ALMA, 
except for the extended rigid rotating massive tori.

\subsection{Expected flux of the HCN molecule}

HCN emission was observed locally in giant molecular clouds and in
nearby galaxies (see i.e. Gao and Salomon 2004 and references
therein). At high redshifts  only gravitationally lensed sources
have been detected, namely H1413+117 (Cloverleaf) at $z$=2.6, IRAS
F10214+4724 at $z$=2.3, and the quasar APM08279+5255 at $z$=3.911
(Solomon et al., 2003; Vanden Bout et al., 2004; Wagg et al.,
2005).

To estimate the flux of the high-$z$ targets discussed
here, we assume that the HCN molecule has the same velocity width
of the CO line, since both emissions are originated within the
same dynamical structure (Wagg et al., 2005). As observed for 
CO lines, we suppose also for HCN luminosity a proportionality with
the molecular hydrogen mass. For normal galaxies a relation
between CO and HCN luminosities is established (Gao \& Solomon,
2004), typically $l^\prime_{\rm HCN}/l^\prime_{\rm CO}\approx 0.1$. 
For most ULIRGs with AGNs this ratio turns to be higher, $>$ 0.2 
(Imanishi et al. 2006).
We take here the value of 0.2, similar to that of Mrk 231
and IRAS 17208-0014, and not far from that of APM08279+5255
(0.34). The HCN line flux is then $l^{\prime}_{\rm
HCN}(1,0)=0.3(X_{\rm HCN}/3.3)^{-1}(X_{\rm CO}/0.8)^{-1}\Sigma_{\rm
g}$, which is similar to the relation $l^{\prime}_{\rm
HCN}(1,0)=\Sigma_{\rm g}/X_{\rm HCN}$, where $X_{\rm HCN}\sim
7~M_\odot({\rm K}\,{\rm km}\,{\rm s}^{-1})$ for warm
gas (Wagg et al. 2005).

%In the same way as CO lines,
%we evaluate the expected flux of the HCN lines by using eq.(3)
%with $X_{\r%m HCN}=3.3~M_\odot(K km s$^{-1}$ pc$^2$)$ and $\nu_{\rm rest}=88\,{\rm GHz%}(1+z)$.
In Fig. 4 we compare the expected HCN line emission from 100 pc scale 
extended massive tori ($M_{\rm BH}=3\times 10^{8}M_{\odot}$, $A=5$, $\alpha=0.5$ and $\beta=1$) as a function of redshift to the ALMA sensitivity limit. 
The dependences of $A$ (or $r_{\rm torus}$), $\alpha$ and $\beta$ are 
same as the case of CO lines ($\S 3.2$).
The detection of HCN lines from massive tori is more difficult 
because of the lower abundance of this molecule with respect to CO. 
At $z\approx 1$ the detection of the central massive torus in this line
would require more than 24 hours of on-source integration in
the highest angular resolution configuration. 
In lower resolution configurations this line is detected at any
redshift, however, the constraint on the model presented in this model
will not be stringent. It will be possible to detect and resolve 
HCN lines for gravitational lensed objects.

\subsection{Which objects are associated to massive tori ?}

We discuss here the best targets associated with massive tori and
their expected number counts. Observationally, the majority of
SMGs is coeval to QSO activities with $z \approx
2$. From their properties (the massive dynamical mass, the high
star formation rate, etc.), SMGs are considered to be progenitors
of present-day elliptical galaxies (e.g., Smail et al. 2004).
Recent ultra-deep X-ray observations have suggested that $> 50\%$
of SMGs have AGNs whose luminosities are relatively lower than
those of coeval QSOs (Alexander et al. 2005a). 
Thus, following our reference model we  expect that massive tori
are already in place in these galaxies and they are feeding the
growth of the central BH with $M_{\rm BH}\approx10^{6}-10^{7}M_{\odot}$ 
(Borys et al. 2005). Recently, Men\'{e}dez-Delmestre et al. (2007) and 
Valiante et al. (2007) observed the mid-infrared spectra of a sample
of SMGs, in which they found that in the rest frame $\sim$ 6-8\,$\mu$m
there is a contribution from a relatively weak AGN continuum emission 
to be added to the contribution from star formation (cfr Fig.3 
in Valiante et al.'s paper). 
Thus, their results stress once again that the most interesting targets 
in the search for tori could be the submm galaxies with 24 $\mu$m counterparts. According to KUM03 (also G04), the dust emission from massive tori in SMGs 
is cooler than that in QSOs because of the higher column density of 
massive tori and relatively weaker AGN activity. 
It will be also worth comparing the rest frame infrared color (e.g., f(25 $\mu{\rm m}$)/f(60 $\mu{\rm m}$)) of SMGs with that of coeval QSOs.
Moreover, Greve et al. (2005), Tacconi et al. (2006) and Gao et al. (2007) 
have reported $\sim $10 CO detected SMGs and $\sim 5$ HCN detected SMGs, 
indicating the presence of 
large masses of dense gas ($\approx 10^{10}M_{\odot}$) 
within $\sim 4\,{\rm kpc}$. 
SMGs are largely present among the Early Universe Molecular Emission 
Lines Galaxies (EMGs) listed by Solomon and Vanden Bout (2005); 
a significant fraction are also lensed objects.

Therefore, SMGs (especially SMGs with AGNs) are good candidates to explore 
a massive torus and to test the radiation drag model. 
We evaluate the expected number count of SMGs having massive tori with the G04 model. 
Note that the G04 model predicts number counts, 
redshift distribution, and AGN accretion rates of
SMGs in very good agreement with observations (G04,
Silva et al. 2005, and Granato et al. 2006).
Fig. 5 shows the
redshift distributions of bright SMGs with different masses of
massive tori. The expected number count of galaxies with $M_{\rm
torus} > 10^{8}M_{\odot}$ is $\sim 100-200$ deg$^{-2}$ at $1 < z
<3$.  By comparing our predictions with ALMA data, it would be
possible test and refine the radiation drag model.

%%% change topic slightly %%%
Lastly, scaled down versions of star forming galaxies
harboring an AGN can be studied in the local Universe.
The narrow line Seyfert galaxies (NLS1s) and ULIRG (especially type I ULIRGs) 
are characterized by a smaller 
BH and a higher mass accretion rate (e.g., Pounds et al. 
1995; Boller et al. 1996; 
Mineshige et al. 2000; Mathur et al. 2000; Collin \& Kawaguchi 2004, 
Hao et al. 2005; Kawakatu et al. 2006; Kawakatu et al. 2007). 
Thus, they would correspond to the early phase of the BH growth.
If massive tori are detected in targets at $z < 0.01$, 
they will be easily resolved with present mm interferometers 
as well. 
Indeed, CO emissions from $ < 100$ pc has already detected 
in nearby NLS1 NGC 4051 (L. Tacconi, private communications). 
By investigating the existences of massive tori in SMGs (spheroids), 
ULIRGs (spheroids and disk galaxies) and NLS1s (disk galaxies), 
it will be possible to clarify whether the BH growth 
process depends on the morphology of host galaxies and redshift. 

\subsection{Velocity mapping on massive tori}

For low-$z$ targets, ALMA can study not only the existence of massive tori, 
but also the sub-pc structures of massive tori, although it is hard to do 
it for high-$z$ objects. 
Thus, we will make the channel maps of molecular lines with ALMA, 
as Davis et al. (2006) carried on it for nearby Seyfert galaxies NGC 3227. 
As a consequence, it is possible to distinguish 
between the circular velocity, $v_{\phi}$ and the velocity dispersion, $\sigma$ in the region above the spatial resolution of ALMA, that is, 
from a few pc to $\sim 100$ pc (e.g., Fig.3 in Wada and Tomisaka 2005). 
If $\sigma$ is comparable to $v_{\rm \phi}$ in massive tori, it means that 
the kinetic energy is dominated by the random motion. 
In order to maintain the vertical structure the continuous energy injection 
should be necessary because the cooling time of molecular clouds via 
collisions is comparable to the orbital period. 
As mentioned in $\S2.2$, the radiation pressure via starburst and/or 
the internal turbulence caused by supernova explosions may be a plausible
origin of its energy source. 
If the scale height of the torus is determined by a balance between 
the vertical components of the centrifugal force due to a central SMBH 
and the vertical velocity dispersion (e.g., Wada \& Norman 2002), 
the ratio $\sigma/v_{\phi}$ reflects the ratio $h_{\rm torus}/r_{\rm torus}$, 
where $h_{\rm torus}$ is the scale height in the body of the massive torus.
Hence, the velocity mapping of massive tori is helpful in providing  
a quantitative estimate of the origin of geometrically thick obscuring torus.

In addition, it is expected that the starburst luminosity in a massive torus is closely related to its vertical structure and the determination of this
through ALMA observations is a fundamental test of this model.
ALMA velocity mapping assesses the velocity dispersion $\sigma$ and 
the scale height $h_{\rm torus}$ for low-$z$ massive tori
and thus it allows to estimate the viscous coefficient of turbulent motion
$\nu_{\rm tub}\sim \sigma h_{\rm torus}$, assuming the largest eddy size 
is comparable to $h_{\rm torus}$. 
Using $\nu_{\rm tub}$, we can estimate the mass 
accretion rate due to the turbulent viscosity $\dot{M}_{\rm BH}\sim 
M_{\rm torus}/t_{\rm vis}$ where $t_{\rm vis}\sim r^{2}/\nu_{\rm tub}$. 
Thus, it will be able to check whether the mass accretion rate via 
the turbulent viscosity is a key element in explaining the AGN activity, because both $M_{\rm torus}$ and  $\nu_{\rm tub}$ are observable. 
In order to get an insight into the physics of a fueling process, 
it would be valuable to derive the radial velocity $v_{\rm r}(r)$, 
which would depend on the mass accretion process (e.g., the gravitational 
torque from a non-axisymmetric potential, the turbulent viscosity, 
the radiation avalanche, etc ).
Therefore, the velocity mapping on massive tori is essential to explore the 
formation of obscuring massive tori and also the fueling process 
from $\sim$100 pc to $\sim$1 pc, which is well known as $^{``}$
missing link" in the AGN fueling problem.

\subsection{Impact on the detection of massive tori for SMBH growth}

In $\S 3.2$, we showed that 100 pc scale extended massive tori 
can be resolved and detected with ALMA. 
Here, we discuss how the detection of massive tori helps in exploring 
the SMBH growth. 
A plausible theoretical process of mass accretion onto a central BH is 
the turbulent viscous drag whose timescale is $t_{\rm vis}\sim r^{2}/
\nu_{\rm tub}$. 
We adopt $\nu_{\rm tub}=R^{-1}_{\rm crit}rv_{\rm \phi}$, where $R_{\rm crit}=100-1000$ and
 $v_{\phi}$ are the critical Reynolds number for the onset of turbulence 
(e.g., Duschl et al. 2000; Burkert \& Silk 2001) and the rotation velocity, 
respectively. 
The viscous time is given by $t_{\rm vis}=R_{\rm crit}
t_{\rm dyn}$, where the dynamical timescale is determined by 
the central BH plus a surrounding massive torus system.
As a result, the mass accretion rate ($\dot{M}_{\rm BH}$) via the viscous drag 
is given as $\dot{M}_{\rm BH}\propto (M_{\rm torus}/M_{\rm BH})^{3/2}[1+(M_{\rm BH}/M_{\rm torus})]^{-1/2}$ (see also eq.(22) in Granato et al. 2004). 
If the massive tori with $M_{\rm torus}/M_{\rm BH}\gg 1$ in SMGs 
are detected with ALMA, the high mass accretion rate
(close to the Eddington or super-Eddington accretion rate) 
would be a common feature of SMGs (especially hard X-ray detected SMGs). 
This statement can be checked comparing the X-ray properties (steep photon index, rapid variability, etc.)
like those seen in NLS1s in future X-ray missions.
 
Moreover, KUM03 (also G04) predicts that the mass ratio 
$M_{\rm torus}/M_{\rm BH}$ can be $\sim 1$ in the QSO phase, 
because the growth of QSO BHs has almost finished and/or
we are seeing the final $e$-folding time (see Fig.1).
It would be then worth comparing the massive tori in QSOs with those expected 
in SMGs with $M_{\rm torus}/M_{\rm BH} \gg 1$ as expected in SMGs.
Thus, this parameter ($M_{\rm torus}/M_{\rm BH}\gg 1$) is a 
key physical conditions for the rapid growth of SMBHs and the appearance 
of QSOs and its evolution tailors the history of the BH growth process.
Considering that the mass of a 100 pc scale dusty torus is much smaller than a central SMBH mass for nearby spheroidal galaxies(e.g., Djorgovski et al. 1991 and Tadhunter et al. 2003 for Cygnus A; Jaffe et al. 1993 and Ferrarese et al. 1996 for NGC 4261; van der Marel et al. 1993 for NGC 7052), 
it would be a natural interpretation that the large amount of gas around central BHs in SMGs accretes
until exhaustion to grow to a SMBH with $\approx 10^{8}-10^{9}M_{\odot}$ 
on timescale of several $10^{8}$ yr. 
Therefore, the detection of massive tori in SMGs 
would imply that the SMBH grows from $\approx 10^{6}M_{\odot}$ to 
$\approx 10^{9}M_{\odot}$ via the gas accretion process mainly and not through the merger of compact objects. 

Finally, we should comment on the case that ALMA detects 
less massive tori in SMGs than what we expect, i.e., $M_{\rm torus}/M_{\rm BH} 
\leq 1$. If this is the case, it might indicate that the growth rate of 
central 
BHs in SMGs is much faster than the mass accretion rate from a host galaxy 
to a massive torus, or some SMGs do not have a potential to evolve into QSOs with $M_{\rm BH}\approx 10^{8}-10^{9}M_{\odot}$. 
Whether or not, we suggest that exploring massive tori would cast light on a SMBH growth process and a formation scenario of QSOs.

\section{Discussions and Conclusions}

Given the sensitivity and angular resolution of ALMA, detailed mapping
of CO emissions of galaxies at $z \sim 2$ will resolve structures
on linear scales of $\sim $ 100 pc and with mass $M_{\rm torus}\approx 
10^{8}-10^{9}$ M$_{\odot}$, the expected size of massive tori around SMBH.
Therefore these capabilities can be used in order to explore the
SMBH growth in SMGs, which are proto-spheroidal galaxies undergoing 
huge star formation, just before the onset of the peak 
of the QSO activity in their centre. 
The results of Alexander et al (2003, 2005a, \& 2005b) show that nuclear X-ray
luminosity $L_X\geq 3\times 10^{43}$ erg${\rm s}^{-1}$ is often associated to SMGs, 
witnessing a stage in which the SMBHs have mass $M_{\rm BH}\sim 10^{6}-
10^{7}M_{\odot}$ (Borys et al. 2005). 
If we can detect massive tori with ALMA within the same population, 
we can definitely conclude that $M_{\rm torus}/M_{\rm BH} \gg 1$ is  
the key physical condition for the rapid growth of SMBHs and 
the appearance of QSOs. 
Moreover, the existence of massive tori in SMGs would imply that 
the SMBH grows from $\approx 10^{6}M_{\odot}$ to $\approx 10^{9}M_{\odot}$ 
via the gas accretion preocess rather than the merger of compact objects.  

In order to assess the observational feasibility, we
estimate the expected number counts of SMGs with massive tori and
check the detectability with the ALMA instrument. ALMA
will be able to resolve and detect high-$J$ CO emissions ($J >4$) 
from massive tori up to $z\approx 2$ at $5\sigma$ level in twelve
hours on-source integration time and a velocity resolution of 25
km/s. 
Lower abundant lines, such as HCN and HCO$^+$ lines,
require much larger integration times at the highest angular
resolution. For even high-$J$ HCN lines at redshift 1, an on-source integration
of at least 24 hours is necessary to detect the massive torus. 
Moreover, we predict the number count
of SMGs with a massive torus (more than $10^{8}M_{\odot}$)
$\approx 100$ deg$^{-2}$ at the redshift $1< z < 3$. 
We stress the
relevance of studying with ALMA  samples of gravitationally lensed
SMGs. Though  lensing introduces the problem of lens model,
nevertheless it can allow a much higher spatial resolution, which
can be a crucial piece of information in inferring the gas
distribution and rotation law 
in the very central regions around the growing SMBHs. 
Studies of lensed and unlensed SMGs would be complementary.

It is also interesting to examine the existences of massive tori
also in nearby AGNs with high mass accretion rate, like 
type I ULIRGs or NLS1s. 
From that, it is possible to reveal whether the BH growth 
process depends on the morphology of host galaxies and redshift.
In addition, determinations of the mass ratio of massive tori and bulges 
in SMGs, ULIRGs and NLS1s would verify directly the radiation drag 
model. This mass ratio is basically determined by the energy conversion efficiency 
of nuclear fusion from hydrogen to helium, 0.007 (Umemura 2001). 
Finally, we should emphasize that the velocity mapping on low-$z$ 
massive tori will be crucial to reveal 
the formation of obscuring massive tori and also the fueling process 
from $\sim$100 pc to $\sim$ 1 pc.

\section*{Acknowledgments} 
We wish to thank the anonymous referee for constructive suggestions 
and fruitful comments. 
We thank M. Umemura for stimulating discussions and 
L. Tacconi and D. Lutz for helpful insights into the topic. 
NK appreciates that M. Umemura allows to edit his 
original figure. We acknowledge the Italian MIUR and INAF 
financial supports.
\label{lastpage}

{}

\clearpage

\begin{table}[t]
\begin{center}
Table. 1  The values of function $H(\alpha,\beta)$ \\[3mm]
{\scriptsize
\begin{tabular}{|l|c|c|c|}
\hline
 & $\beta=0$ & $\beta=1$ & $\beta=2$ \\
\hline
$\alpha=-1$ & 2 & 1 & 20 (7) \\
\hline
$\alpha=0$  & 1  & 1 &  1    \\
\hline
$\alpha=0.5$ & 0.8 & 0.7$^a$ & 0.4 (2) \\
\hline
\end{tabular}
}
\noindent
\end{center}
{\scriptsize Note.--- The values in parenthesis are for 
20 pc compact massive tori. The others correspond to the 
case of 100 pc extended massive tori.\\
The symbol $a$ corresponds to the fiducial case in this paper.
}
\end{table}

\newpage

\begin{figure}
\begin{center}
%\textwidth=7.5cm
%\plotone{fig1.eps}
\includegraphics[height=7cm,clip]{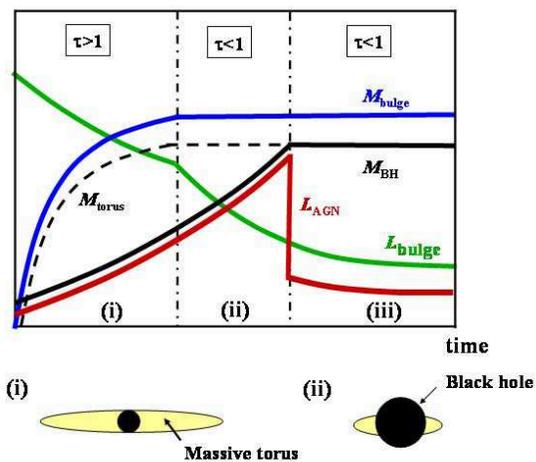}
\end{center}
%\vspace{-0.5cm}
\caption{
{
A schematic sketch of the co-evolution of a SMBH and a galactic bulge 
(We edit the original figure (Fig. 1) in Umemura 2001.).
The abscissa is time and the ordinate is arbitrary .
%$M_{\rm bulge}$ is the mass of the stellar component in the bulge.
%$M_{\rm torus}$ is the mass of the massive torus around a central BH.
%$M_{\rm BH}$ is the mass of the supermassive BH. $L_{\rm bulge}$ and
%$L_{\rm AGN}$ are the bulge luminosity and the AGN luminosity via mass acc%retion, respectively.
$\tau$ is the optical depth of bulges.
The phase (i) denotes the early phase of a QSO (SMG phase). The QSO phase corresponds to the end of the phase (ii).
In the phase (i), a massive torus around a smaller BH is predicted.
}
}
\end{figure}

\newpage
\begin{figure}
\begin{center}
%\textwidth=7.5cm
%\plotone{fig1.eps}
\includegraphics[height=6cm,clip]{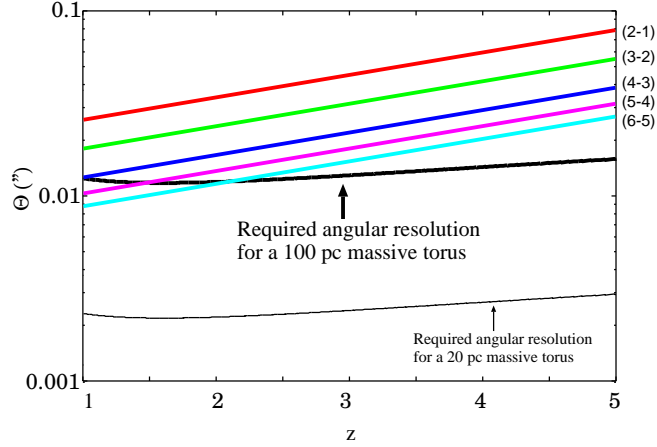}
\end{center}
%\vspace{-0.5cm}
\caption{
{
The required spatial resolution (arcsec) for a 20 pc scale (thin horizontal line) and  a 100 pc scale (thick horizontal line) massive torus 
against redshift $z$. 
The five lines represent the possible spatial resolution for CO lines in 
the transition level ($J, J-1$) with the ALMA instrument.
}
}
\end{figure}

\newpage
\begin{figure}
\begin{center}
%\textwidth=7.5cm
%\plotone{fig1.eps}
\includegraphics[height=6cm,clip]{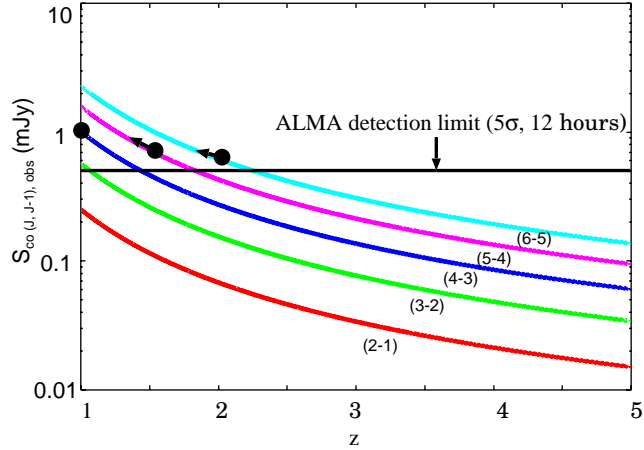}
\end{center}
%\vspace{-0.5cm}
\caption{
{
The expected CO emission (mJy) from massive tori against redshift $z$ in the fiducial case ($M_{\rm torus}=3\times 10^{8}M_{\odot}$, $A=5$, $\alpha=0.5$ and $\beta=1$). The five lines denote the CO flux density $S_{\rm CO}(J,J-1)_{\rm obs}
$ at the transition level $(J, J-1)$ at the rest frame. The horizontal line represents the ALMA detection limit 0.5$\,$mJy (50 antennas) requested $5\sigma$ in 25 km/s channel in 12 hours of on-source integration time. At the left side of filled circle, each line can be resolved spatially. 
%In the case of $M_{\rm torus}=10^{9}M_{\odot}$,  the expected CO emission incre%ases with by factor 2 as the arrow shows.
}
}
\end{figure}

\newpage
\begin{figure}
\begin{center}
%%\textwidth=7.5cm
%%\plotone{}
\includegraphics[height=6cm,clip]{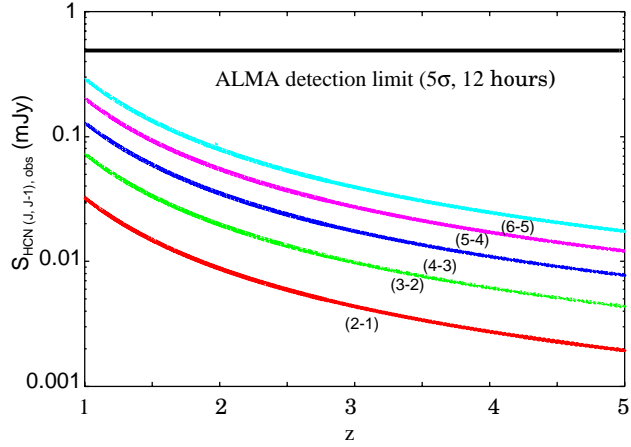}
\end{center}
%\vspace{-0.5cm}
\caption{
The expected HCN emission (mJy) from massive tori against
redshift $z$ in the fiducial case 
($M_{\rm torus}=3\times 10^{8}M_{\odot}$, $A=5$, $\alpha=0.5$ and $\beta=1$).
The lines are same as Fig. 3.
}
%\label{fig:1}
\end{figure}

\begin{figure}
\begin{center}
%%\textwidth=7.5cm
%%\plotone{fig1.eps}
\includegraphics[height=8cm,clip]{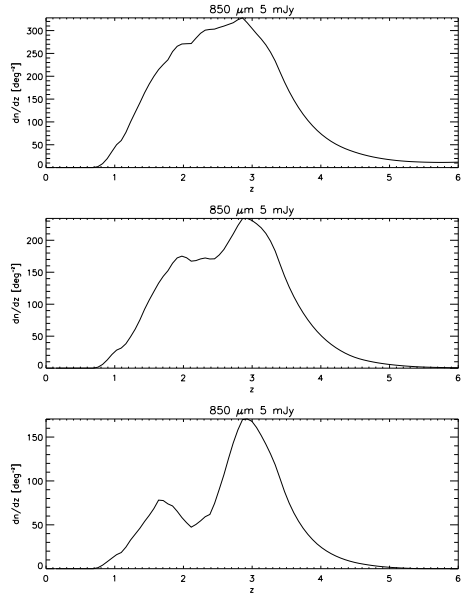}
\end{center}
%\vspace*{-1.0cm}
\caption{
Number counts of SMGs with massive tori $dn/dz\,[{\rm deg}^{-2}]$ against redshit $z$. Top panel shows the number counts of all SMGs. The middle one denotes $dn/dz\,[{\rm deg}^{-2}]$ with $M_{\rm torus} > 10^{8}M_{\odot}$. The bottom one shows the the number counts of galaxies with massive tori $M_{\rm torus} > 3\times 10^{8}M_{\odot}$.
}
%\label{fig:1}
\end{figure}

%\label{lastpage}
\end{document}